\documentstyle[12pt]{article}

\begin{document}

\title {\bf \Large Dirac Equation in Standard Cosmological Models}

\author{M. Sharif \thanks{e-mail: hasharif@yahoo.com}
\\ Department of Mathematics, University of the Punjab,\\ Quaid-e-Azam
Campus Lahore-54590, PAKISTAN.}

\date{}

\maketitle

\begin{abstract}
The time equation associated to the Dirac Equation (DE) is studied
for the radiation-dominated Friedmann-Robertson-Walker (FRW)
universe. The results are analysed for small and large values of
time. We also incorporate the corrections of the paper studied by
Zecca [1] for the matter-dominated FRW universe.
\end{abstract}
{\bf PACS 03.65.Pm-Relativistic wave equations.
}

\newpage

\section{Introduction}

It is of cosmological interest to consider a universe of
incoherent radiation. The primordial radiation played a dominant
role in the early universe for the times smaller than the times of
re-combination. Zecca [1] has discussed the time equation
associated to the DE in the FRW spacetime in the case of Standard
Cosmology. He confined his paper to the study of matter-dominated
cosmological model only. This paper is dedicated to a further
study of the time equation associated to the DE of a
radiation-dominated cosmological model. Also, we will incorporate
the corrections in the expansion terms of Eqs.(14) and (17) in the
Zecca's paper [1].

It is well known that the DE can be written in General Relativity
in terms of covariant derivatives and generalised Pauli matrices.
In the context of the Newmann-Penrose formalism [2], the DE can
further be expressed in terms of directional derivatives and spin
coefficients. The DE, formulated by the spinorial formalism of
Newmann and Penrose [3], is given by

\begin{equation}
\left.
\begin{array}{l}
\nabla_{AA^\prime}P^A+\iota\mu_*\bar Q_{A^\prime}=0,\\
\nabla_{AA^\prime}Q^A+\iota\mu_*\bar P_{A^\prime}=0,
\end{array}
\right \}
\end{equation}
where $\nabla_{AA^\prime}$ are the covariant spinor derivatives
and $\mu_*\sqrt{2}$ is the mass of the particle [4]. If we
correspond $\phi\leftrightarrow(P,Q),\quad
\psi\leftrightarrow(U,V)$, we can define the spinor

\begin{equation}
J^{AA^\prime}(\phi,\psi)=P^A\bar U^{A^\prime}+V^A\bar Q^{A^\prime}
\end{equation}
This is divergence free, i.e.,
$\nabla_{AA^\prime}J^{AA^\prime}\equiv\nabla_\alpha J^\alpha=0$
due to Eq.(1). We can define inner product between the solutions
of the DE by setting [5]

\begin{equation}
(\phi,\psi)\equiv\int_\Sigma J_\alpha(\phi,\psi)(-g\Sigma(x))^{1/2}n^\alpha d\Sigma
\end{equation}

\begin{equation}
=\int_{t=t_0}d_3x(-g_{t_0})^{1/2}\sigma_{AA^\prime}^tJ^{AA^\prime}(\phi,\psi)
\end{equation}

\begin{equation}
=\frac{1}{2}\int_{t=t_0}d_3x(-g_{t_0})^{1/2}(P^0\bar{U^0}+P^1\bar{U^1}
+V^0\bar{Q^0}+V^1\bar{Q^1}).
\end{equation}
The Eq.(3) is independent of the spacelike Cauchy hypersurface
$\Sigma$ of volume element $d\Sigma, \quad n^\alpha$ is a future
directed unit vector orthogonal to $d\Sigma, \quad
\sigma_{AA^\prime}^t$ is the generalised Pauli matrix and has the
form $ \sigma_{AA^\prime}^t=\frac{1}{2}\left( \begin{array}{cc}
1&0\\0&1\end{array} \right)$.

The separation of the angular part in Eq.(1) is performed by a
standard separation method which gives the Teukolski-like equation
for spin(1/2) field admitting explicit solution [6]. The surviving
coupled equations in the time and radial variables are separated
by writing the unknown wave function in terms of a known
particular solution. These equations can be expressed in the
conformal time parameter $\tau$ and in the spatial parameter s
defined by

\begin{equation}
\tau(t)=\int_0^t\frac{dt^\prime}{R(t^\prime)},\quad
s(r)=\int_0^r\frac{dr^\prime}{\sqrt{1-ar^{\prime2}}},\quad
(a=0,\pm1).
\end{equation}
The complete spinorial solution of Eq.(1) has been found and is of the form

\begin{equation}
\left.
\begin{array}{l}
P^i=\frac{S_{lm}^{(i+1)}}{2rR(t)}H(r,t)\{(-1)^iA_{lk}(s)T_k(\tau)+[\frac{2\lambda}{r}A_{lk}
(s)-A_{lk}^\prime(s)]\int_0^\tau d\hat\tau T_k\},\\
\bar Q_i=-\frac{S_{lm}^{(2-i)}}{2rR(t)}H(r,t)\{A_{lk}(s)T_k(\tau)+(-1)^i[\frac{2\lambda}
{r}A_{lk}(s)-A_{lk}^\prime(s)]\int_0^\tau d\hat\tau T_k\},
\end{array}
\right\}
\end{equation}
where $i=0,1$. The angular functions appearing in Eq.(7) are of
the form $S_{lm}^{(j)}=S_{lm}^{(j)}(\theta,\phi)'s(j=1,2)$, where
$S_{lm}^{(j)}(\theta,\phi)$ are solutions of an eigen-value
problem originated by the solution of the angular part of Eq.(1).
The $A_{lk}(s)'s$ are the solutions of the angular and radial
equations, respectively whose explicit expressions can be found in
[7]; $\lambda, k$ are separation constants such that
$\lambda^2=(l+1)^2, l=0,1,2,...$ for $m=0; \lambda^2=(l+1/2)^2, l=
|m|,|m|+1,|m|+2,...$ for $m=0,\pm1,\pm2,\pm3,...$. The angular
functions are assumed to satisfy the normalization condition

\begin{equation}
\int d\Omega S_{lm}^{(j)}(\theta,\phi)S_{l^\prime
m^\prime}^{(j)}(\theta,\phi)
=\delta_{ll^\prime}\delta_{mm^\prime}.
\end{equation}
The function $H$ is connected to a particular integral of Eq.(1) and has the form

\begin{equation}
H(r,t)=\frac{1}{R^{1/2}(t)}\left (\frac{1+\sqrt{1-ar^2}}{r}\right
)^\lambda exp[\iota\mu_*\sqrt2t],\quad (a=0,\pm1).
\end{equation}

The function $T_k(\tau)$ is a solution of the separated time
equation in the conformal time parameter $\tau$ and satisfies [7]
the equation

\begin{equation}
T^{\prime\prime}+2\sqrt2\iota\mu_*R(\tau)T^\prime+(2\iota\sqrt2\mu_*R^\prime(\tau)+k^2)T=0,
\end{equation}
where prime denotes differentiation with respect to $\tau$. The
solution of Eq.(10) depends on the dynamical evolution of the
cosmological background and this satisfies a property similar to
the Wronskian property of the scalar field time equation in FRW
spacetime [8,9]. By using a first formal integration of Eq.(10),
one can easily show that any solution $T_k(\tau)$ of Eq.(10)
satisfies the relation

\begin{equation}
|T_k(\tau)|^2+k^2\left|\int_0^\tau
T_k(\hat\tau)d\hat\tau\right|^2=constant=1.
\end{equation}
This normalization is a necessary requirement of a second
quantization of the Dirac field. It is noted that Eq.(10) can be
solved in particular simple cases such as that of static universe
or of the neutrino case. However, it seems difficult to give the
solution of Eq.(10) for a general cosmological evolution.

The layout of this paper is as follows. In section two, we analyse
the matter-dominated FRW universe to incorporate the corrections
studied by Zecca [1]. In the next section we shall extend this
procedure to study the solutions for the radiation-filled
cosmological model. Section four is devoted to discussion of the
results.

\section{Solutions in the Matter-Dominated Cosmological Model}

It is well known that the FRW spacetime whose metric

\begin{equation}
ds^2=dt^2-R^2(t)\left[\frac{dr^2}{1-ar^2}+r^2(d\theta^2+\sin^2\theta
d\phi^2)\right],\quad(a=0,\pm1)
\end{equation}
is the base of the standard model [10,11]. This is the natural
context for the formulation of the Einstein Field Equation for a
general isotropic and homogeneous universe. We shall first
consider the matter-dominated universe for the purpose of
correction in [1]. For matter-dominated universe, $R(t)$ in
parametric form is given as

\begin{equation}
\left.
\begin{array}{l}
R(t)=Et^{2/3}, \quad(a=0),\\
R(t)=A(\cosh\psi-1),\quad t=B(\sinh\psi-\psi),\quad\psi\ge0,\quad(a=-1),\\
R(t)=C(1-\cos\theta),\quad t=D(\theta-\sin\theta),\quad 0\le\theta\le 2\pi,\quad(a=1).
\end{array}
\right\}
\end{equation}
In terms of the conformal time parameter $\tau$ defined in Eq.(6) we obtain

\begin{equation}
\left.
\begin{array}{l}
R(\tau)=E^3\tau^2/9,\quad(a=0),\\
R(\tau)=A[\cosh(\tau A/B)-1],\quad(a=-1),\\
R(\tau)=C[1-\cos(\tau C/D)],\quad(a=1).
\end{array}
\right\}
\end{equation}
If we use Eq.(14) in Eq.(10) they all give a linear differential
equation with finite regular point of the form
\begin{equation}
T^{\prime\prime}+\left(\sum_{n=0}^\infty p_n\tau^n\right)T^\prime+\left(\sum_{n=0}^\infty
q_n\tau^n\right)T=0.
\end{equation}
By using a standard method [12], we can set

\begin{equation}
T=\sum_{n=0}^\infty c_n\tau^n,
\end{equation}
where the coefficients $c_n$ can be found by the recurrence relation

\begin{equation}
(n+2)(n+1)c_{n+2}+\sum_{j=0}^\infty p_j(n+1-j)c_{n+1-j}+\sum_{j=0}^\infty q_jc_{n-j}=0,
\quad n=0,1,2,...
\end{equation}
We shall always take $c_1=0$ due to our interest only in the time
integral.

For the flat model $a=0$, the recurrence relation (14) in [1] will
become

\begin{equation}
\begin{array}{l}
T=c_0 \{1-\frac{\tau^2}{2!}k^2-\frac{\tau^3}{3!}2\alpha
E^3/9+\frac{\tau^4}{4!}k^4 +\frac{\tau^5}{5!}14\alpha k^2 E^3/9
\\
\qquad +\frac{\tau^6}{6!}[40(\alpha
E^3/9)^2-k^6]-\frac{\tau^7}{7!}44\alpha
E^3k^4/9+\frac{\tau^8}{8!}[k^8-628k^2(\alpha E^3/9)^2]\\ \qquad
\qquad \qquad \qquad +\frac{\tau^9}{9!}[100\alpha
E^3k^6/9-2240(\alpha E^3/9)^3]+... \},
\end{array}
\end{equation}
where $\alpha=2\sqrt2\iota\mu_*$.

For the open model $a=-1$, the recurrence relation (17) in [1]
will take the form

\begin{equation}
\begin{array}{l}
T=c_0\{1-\frac{\tau^2}{2!}k^2-\frac{\tau^3}{3!}\alpha
A^3/B^2+\frac{\tau^4}{4!}k^4 +\frac{\tau^5}{5!}[7\alpha
A^3k^2/B^2-\alpha A^5/B^4]\\ \quad +\frac{\tau^6}{6!}[10(\alpha
A^3/B^2)^2-k^6]+\frac{\tau^7}{7!}(16\alpha A^5k^2/B^4-22\alpha
A^3k^4/B^2-\alpha A^7/B^6)\\ \qquad \qquad \qquad
\qquad+\frac{\tau^8}{8!}[-157(\alpha A^3/B^2)^2k^2+56\alpha^2
A^8/B^6+k^8]+...\}.
\end{array}
\end{equation}
By combining Eqs.(10),(13) and (14) one has $T\cong R^{-1}\cong
t^{-2/3}\cong\tau^{-2}$ for $\tau\gg1$. Thus by using Eqs.(7),(9)
and (18), the time dependence of $H$ will take the form

\begin{equation}
t^{-1/3}exp[\iota\mu_*\sqrt2t],\quad(t\gg1,a=0).
\end{equation}

\section{Solutions in Radiation-Dominated Cosmological Model}

In this section we shall extend the above procedure to solve
Eq.(10) for radiation-filled FRW universe model. For
radiation-dominated FRW spacetimes, the scale factor $R$ is given
as

\begin{equation}
\left.
\begin{array}{l}
R(t)=Et^{1/2},\quad (a=0),\\
R(t)=A\sinh\psi,\quad t=B(\cosh\psi-1),\quad \psi\ge0,\quad(a=-1),\\
R(t)=C\sin\theta,\quad t=D(1-\cos\theta),\quad 0\le\theta\le2\pi,\quad(a=1).
\end{array}
\right\}
\end{equation}
We can write these values of scale factor in terms of the
conformal time parameter $\tau$ as

\begin{equation}
\left.
\begin{array}{l}
R(\tau)=E^2\tau/2,\quad (a=0),\\
R(\tau)=A\sinh(\tau A/B),\quad(a=-1),\\
R(\tau)=C\sin(\tau C/D),\quad(a=1).
\end{array}
\right\}
\end{equation}
\begin{description}
\item {\bf (i) The Flat Model a=0}
\end{description}

By using Eqs.(10), (15) and the 1st equation in (22), we can have

\begin{equation}
\left.
\begin{array}{l}
p_0=0,\quad p_1=\alpha E^2/2,\quad p_n=0,\quad n\ne1,\\
q_0=\alpha E^2/2+k^2,\quad q_n=0,\quad n\ge1.
\end{array}
\right\}
\end{equation}
With the help of Eq.(23) in the recurrence relation (17), we can
find all values of c. Replacing these values of the c's in
Eq.(16), we get

\begin{equation}
\begin{array}{l}
T=c_0\{1-\frac{\tau^2}{2!}(\alpha
E^2/2+k^2)+\frac{\tau^4}{4!}(3\alpha E^2/2+k^2)(\alpha
E^2/2+k^2)\\ \qquad-\frac{\tau^6}{6!}(5\alpha E^2/2+k^2)(3\alpha
E^2/2+k^2)(\alpha E^2/2+k^2)\\ \qquad+\frac{\tau^8}{8!}(7\alpha
E^2/2+k^2)(5\alpha E^2/2+k^2)(3\alpha E^2/2+k^2)(\alpha
E^2/2+k^2)+...\}.
\end{array}
\end{equation}
\begin{description}
\item {\bf (ii) The Open Model a=-1}
\end{description}

From Eqs.(10), (15) and the second in Eq.(22) with the use of the
series expansions of $\cosh x, \sinh x$, we can make the
identifications

\begin{equation}
\left.
\begin{array}{l}
p_0=0,\quad p_{2n-1}=\frac{\alpha A}{(2n-1)!}(A/B)^{2n-1},\quad n\ge1,\\
q_0=\alpha A^2/B+k^2,\quad q_{2n}=\frac{\alpha A^2}{B(2n)!}(A/B)^{2n},\quad n\ge1.
\end{array}
\right\}
\end{equation}
Making use of these values in the recurrence relation (17) to find
c's and then Eq.(16) yields

\begin{equation}
\begin{array}{l}
T=c_0\{1-\frac{\tau^2}{2!}(\alpha
A^2/B+k^2)+\frac{\tau^4}{4!}[(3\alpha A^2/B+k^2)(\alpha
A^2/B+k^2)\\ \qquad \qquad-\alpha
(A^2/B)(A/B)^2]-\frac{\tau^6}{6!}[(5\alpha A^2/B+k^2)\{(3\alpha
A^2/B+k^2)\\ \qquad \qquad(\alpha A^2/B+k^2)-\alpha
(A^2/B)(A/B)^2\}-10\alpha (A^2/B)(A/B)^2\\ \qquad \qquad(\alpha
A^2/B+k^2) +\alpha (A^2/B)(A/B)^4]+\frac{\tau^8}{8!}[(7\alpha
A^2/B+k^2)\\ \qquad \qquad \{(5\alpha A^2/B+k^2)\{(3\alpha
A^2/B+k^2)(\alpha A^2/B+k^2)\\ \qquad \qquad-\alpha
(A^2/B)(A/B)^2)\}-10 \alpha (A^2/B)(A/B)^2(\alpha A^2/B+k^2)\\
\qquad \qquad+\alpha (A^2/B)(A/B)^4\}-35 \alpha (A^2/B)(A/B)^2
\{(3\alpha A^2/B+k^2)\\ \qquad \qquad(\alpha A^2/B+k^2)-\alpha
(A^2/B)(A/B)^2)\}+3\alpha (A^2/B)(A/B)^4\\ \qquad \qquad(\alpha
A^2/B+k^2)-\alpha (A^2/B)(A/B)^6]+...\}.
\end{array}
\end{equation}
\begin{description}
\item {\bf (iii) The Closed Model a=1}
\end{description}

The closed model can be obtained from the open model $(a=-1)$ by
replacing $A$ by $\iota C$ and $B$ by $-D$. Using the
normalization Eq.(11), we can have $c_0=1$ in every case.

\section{Discussion of the Reults}

We see from Eqs.(24) and (26) that when $\tau$ tends to zero
$T(\tau)$ approaches to 1. This implies that the complete
spinorial solution (7) becomes independent of the conformal time
parameter $\tau$ for $\tau \to 0$. When we combine
Eqs.(7),(9),(21),(22),(24) and (26), we obtain the same behaviour
in every case for $t<<1$.

Consider the case of a cosmological model with $a=0$. For large
$t, R\cong t^{1/2}$ hence $R\cong\tau$ so that the asymptotic
behaviour of the solutions of Eq.(10) are now approximated by
$T\cong\tau^{-1}\cong t^{-1/2}$ so that $H$ has the time
dependence

\begin{equation}
t^{-1/4}exp[\iota\mu_*\sqrt2t],\quad(t\gg1,a=0)
\end{equation}
Similarly, for the open $(a=-1)$ radiation-dominated universe we
have $R\cong t$ and hence $R\cong exp(\tau A/B)$ for large $t$.
Thus the asymptotic behaviour of the solutions of Eq.(10) is given
by $T\cong exp(-\tau A/B)\cong t^{-1}$ and for $H$ the time
dependence is of the form

\begin{equation}
t^{-1/2}exp[\iota\mu_*\sqrt2t],\quad(t\gg1,a=-1)
\end{equation}

\begin{description}
\item {\bf Acknowledgments}
\end{description}

The author would like to thank Prof. Chul H. Lee for his
hospitality at the Department of Physics and Korea Scientific and
Engineering Foundation (KOSEF) for postdoc fellowship at Hanyang
University Seoul, KOREA.

\vspace{2cm}

{\bf \large References}

\begin{description}

\item{[1]} Zecca, A.: Nuovo Cimento {\bf B113}(1998)915.

\item{[2]} Montaldi, E. and Zecca, A.: Int. J. Theor. Phys. {\bf 37}(1998)995.

\item{[3]} Newman, E. and Penrose, R.: J. Math. Phys. {\bf 3}(1962)566.

\item{[4]} Chandrasekhar, S.: {\it The Mathematical Theory of Black Holes}
(Oxford University Press, 1983).

\item{[5]} Hawking, S.W. and Ellis, G.F.R.: {\it The Large Scale Structure of
Spacetime} (Cambridge University Press, 1973).

\item{[6]} Montaldi, E. and Zecca, A.: Int. J. Theor. Phys. {\bf
33}(1994)1053.

\item{[7]} Zecca, A.: J. Math. Phys. {\bf 37}(1996)874.

\item{[8]} Birrell, N.D. and Davies, P.C.W.: {\it Quantum Fields in Curved
Spacetime} (Cambridge University Press, 1982).

\item{[9]} Zecca, A.: Int. J. Theor. Phys. {\bf 36}(1997)1387.

\item{[10]} Misner, C.W., Thorne, K.S. and Wheeler, J.A.: {\it Gravitation}
(W.H. Freeman San Francisco, 1973).

\item{[11]} Peebles, P.J.E.: {\it Principles of Physical Cosmology} (Princeton
University Press, 1993);

Kolb, E.W. and Turner, M.S.: {\it The Early Universe} (Addison Wesley, 1990).

\item{[12]} Arfken, G.B. and Weber, H.J.: {\it Mathematical Methods for
Physicists} (Academic Press New York, 1995).

\end{description}

\end{document}